\documentclass[12pt]{article}
\usepackage{pic04}
\usepackage{hyperref}
\usepackage{graphicx}

\begin{document}

\title{\bf QUARKONIUM PRODUCTION AND DECAY}
\author{
Richard S. Galik       \\
{\em Cornell University, Ithaca, NY 14853}}
%

\maketitle

%

%
%
%
\vspace{4.5cm}
%

\baselineskip=14.5pt
\begin{abstract}
With new data sets, recently completed analyses and 
renewed interest, there has been significant progress in addressing
existing questions about quarkonia production and decay ...
but also new questions and new confrontation with theory.
Some highlights include the firm establishment of the 
$\eta_{c}^{\prime}$ and of the ``missing'' $\psi^{\prime}$
decays to $0^- 1^-$ final states, improved information on the nature
of the $\psi$(3770), and the observation of a new, puzzling charmonium-like
state at 3872 MeV.
\end{abstract}
\newpage

\baselineskip=17pt

\section{\bf Introduction}
\label{sec:introduction}

\subsection{Why study Quarkonia?}

Quarkonia, bound states of a quark $q$ and its
anti-quark $\overline{q}$, are the QCD equivalents
of positronium (bound $e^{+}e^{-}$) in QED.  Just as we
have learned much about QED from the spectra of
positronium, so too we probe the strong force with
$q\overline{q}$ bound states.  Quarkonia form the
simplest, most symmetric strongly interacting system
with only two constituents (unlike baryons) and
those being identical (unlike mesons with ``open'' flavor).

The QCD potential is much richer than that of QED.  In the
simplest formulation, the so-called ``Cornell'' potential,
we have:
\begin{equation}
\label{eqn:Cornell}
V(r) = -\frac{4}{3}\cdot\frac{\alpha_{s}}{r} + k \cdot r
\end{equation}
\noindent
The first term mimics the single-photon exchange term
in QED and corresponds to one-gluon exchange in QCD,
dominating at short distances ($< 0.1 $fm) 
with large momentum transfers.
This is the asymptotically ``free'' regime which has 
$\alpha_{s}$ small, making calculations perturbative
(PQCD).
The second term, important at larger distance scales ($> 1$ fm),
leads to ``confinement'' and is in a regime where
$\alpha_{s}$ is large, making calculations non-perturbative.

In this presentation I will discuss only {\it heavy} quarkonia,
$Q\overline{Q}$, with $Q = c$ (charmonium)  
or $b$ (bottomonium).\footnote{Toponia would be wonderful to study,
but the with $m_{t} > m_{W}$ the top quark is not
sufficiently stable to form hadrons before decaying.}

A big advantage of such $Q\overline{Q}$ systems is that
they are not very relativistic:
$\beta^{2}(c\overline{c}) \sim 0.25$ and
$\beta^{2}(b\overline{b}) \sim 0.08$.  Therefore non-relativistic QCD
(NRQCD) can be used to good approximation, greatly simplifying
calculations. 
Of particular note\footnote{See the contribution of
Matt Wingate to this conference} is the fact that
lattice QCD (LQCD) is making important advances in
predictive power for these systems.
Further, the physical sizes of $Q\overline{Q}$
states span the range of ``free'' and ``confined'', allowing
both aspects of QCD to be studied.

While this presentation concentrates on spectra and decays,
one cannot ignore the importance of production in that much is
to be learned from the formation of such $Q\overline{Q}$ 
systems from (virtual) photons and gluons.

\begin{figure}[htbp]
  \centerline{\hbox{
   \includegraphics[width=13.0cm]{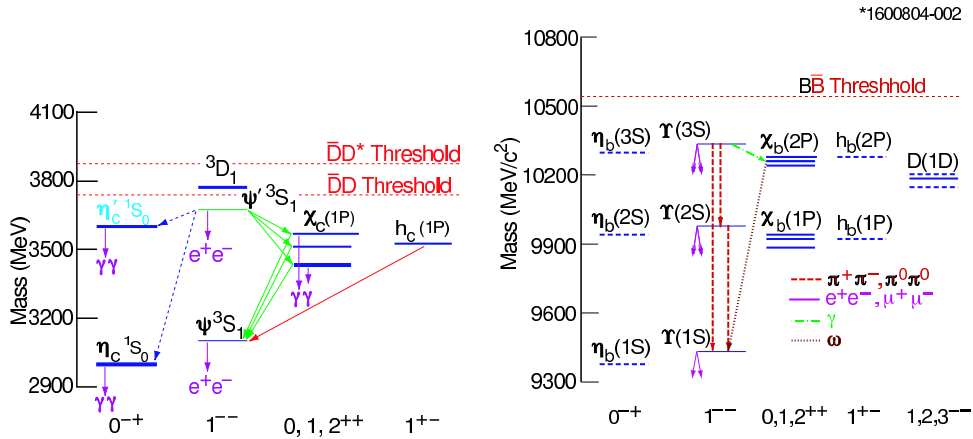}
    } 
   }  
  \caption{\it On the left (right) is the charmonium (bottomonium)
spectrum below 
and near open flavor
threshold.  Shown for
$c\overline{c}$ are
di-lepton and two-photon decays, which, by t-reversal, are typical
production mechanisms. 
The singlet $h_{c}$ remains unconfirmed.  
Shown for
$b\overline{b}$ are some of the
transitions and decays with  $\Upsilon$(3S) as parent,
indicating the diversity of modes and the large amount of
available energy release, $Q$.  None of the five singlet 
$\eta_{b}$ or
$h_{b}$ states have yet to be observed.}
 \label{fig:charmbottom} 
\end{figure}

The spectra for charmonia and bottomonia are shown in
Figure~\ref{fig:charmbottom}, 
with some typical transitions
indicated.  Note that the QCD wells for such
states are much deeper than the corresponding case for
positronium, leading to a much richer spectrum, 
involving many {\it hadronic} transitions in
addition to those associated with photon emission.
In particular, the $b\overline{b}$ spectrum has three
radial excitations below open bottom threshold and enough
energy splitting to allow multi-pion or $\omega$ emission
in transitions among states.
%
%

\subsection{Disclaimers and Previous Reviews}
This document is
a faithful
replication of the PIC04 presentation 
being a non-exhaustive {\it selection}
of topics.
Discoveries
or updates made after PIC04 will not be included; 
this is
therefore a snapshot of the field
on 27 June 2004.  One exception 
is that in the references
I 
give the {\it current} status of the work. 

Of course this review builds on previous efforts,
of which two good examples are 
the PIC03 contribution
of Mahlke-Kr\"uger \cite{HMK} and that from the 2003
Lepton-Photon conference by Skwarnicki\cite{TS}.  There
is also a comprehensive review of quarkonia underway
by the Quarkonium Working Group\cite{QWG}; this CERN
Yellow Report is to be completed by September 2004.

\section{\bf News on the $Q\overline{Q}$ Spin-Singlets}

These states have the spins of $Q$ and $\overline{Q}$
in an anti-symmetric state, so that $S = 0$ and $J = L$.
The states with 
$J^{PC} = 0^{-+}$ (pseudoscalars) are called ``$\eta_{Q}$'' 
and those with 
$J^{PC} = 1^{+-}$ are called ``$h_{Q}$''.

For 
$b\overline{b}$
there is no news on these singlets; the CLEO
limits\cite{TS} from 2003 have been slightly
updated\cite{QWG} but none of the five have been observed
as yet.
In the $c\overline{c}$ sector these have been some new measurements
of the mass, width and decays of the ground state of charmonium,
the $\eta_{c}$, but no startling developments. The lone singlet
$P$ state of charmonium, the $h_{c}$, remains elusive, with
active programs in $p\overline{p}$ production, $\psi^{\prime}$
decay and $B \to (c\overline{s}) K$ (and maybe others!) all in the 
hunt.

The biggest news from the spin singlets is the (re)discovery of the
$\eta_{c}^{\prime}$.  This had been previously reported 
by the Crystal Ball experiment \cite{CBetacpr} in the M1 radiative
transition $\psi^{\prime} \to \gamma\eta_{c}^{\prime}$. But CLEO,
with similar sensitivity to the Crystal Ball,\footnote
{For example, CLEO {\it does} see the hindered transition to
$\gamma\eta_{c}$ at $8\sigma$ significance.} does not confirm
that observation.\cite{TS,Hajime} 

\begin{figure}[htbp]
  \centerline{\hbox{ \hspace{0.2cm}
    \includegraphics[width=6.2cm, height = 4cm]{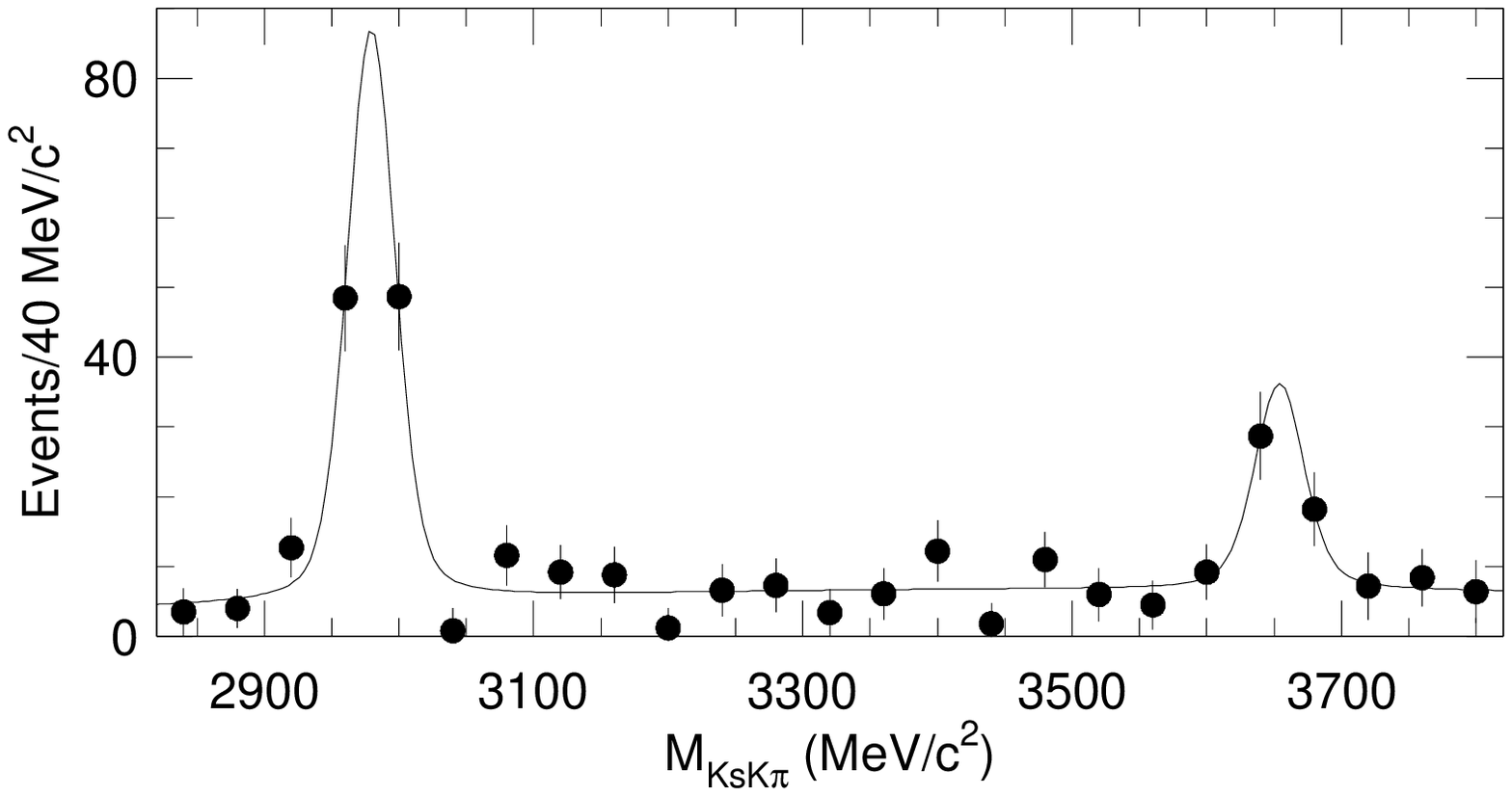}
    \hspace{0.3cm}
    \includegraphics[width=5.8cm]{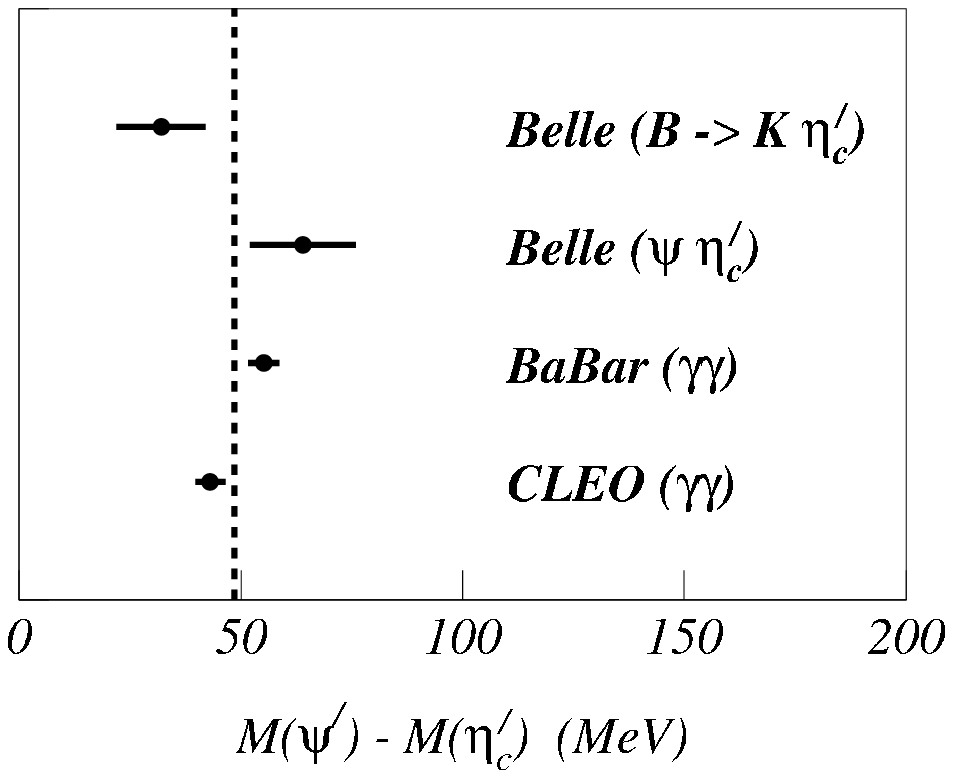}
    } 
   }  
  \caption{\it First observations of the $\eta_{c}^{\prime}$.
On the left is the Belle spectrum in $B \to K (K_{S}^{0}K\pi)$, 
showing the well-known $\eta_{c}$, 
and observation of the $\eta_{c}^{\prime}$
at 3654 MeV.  To the
right is a summary of the four measurements by Belle, BaBar and CLEO.
The weighted average of the hyperfine splitting to the $\psi^{\prime}$
is $49 \pm 2$ MeV.}
 \label{fig:etacpr} 
\end{figure}

But four {\it new} observations of this radially excited singlet
have now been {\it published}.  First, Belle
presented a very clean signal (see 
Fig.~\ref{fig:etacpr}) in $B\to\eta_{c}^{\prime} K$ decays \cite{Belle1};
they also observed the state in continuum production of double
charmonium, $e^{+}e^{-}\to J/\psi~\eta_{c}^{\prime}$ \cite{Belle2}.
This was followed by both CLEO\cite{CLEOetacpr} and BaBar\cite{BaBaretacpr}
publishing their measurements of the $\eta_{c}^{\prime}$ in two-photon
fusion.  

The four new measurements of the mass of the $\eta_{c}^{\prime}$ are shown
in the right portion of Fig.~\ref{fig:etacpr}, presented\cite{Zaza} as the
hyperfine mass splitting between it and the spin-triplet $\psi^{\prime}$.
The weighted average for this splitting is 
\begin{equation}
\Delta_{hf}^{\prime} 
= m(\psi^{\prime}) - m(\eta_{c}^{\prime}) = (49 \pm 2 ) {\rm MeV} 
\end{equation}
which is roughly {\it half} the magnitude for
this splitting based on the older Crystal Ball measurement of
$m(\eta_{c}^{\prime})$,
and which is to be compared with 
$\Delta_{hf} = m(J/\psi) - m(\eta_{c})  = (117 \pm 2 )$ MeV.
\footnote{The four results have sufficient spread that using the
PDG\cite{PDG2002} 
prescription for scaling uncertainties leads to $49 \pm 4$ MeV
for this hyperfine splitting.}

The splittings $\Delta_{hf}$ and $\Delta_{hf}^{\prime}$ are of 
some interest because, while the $\eta_{c}$ and $J/\psi$ have
small radius and are rather deep in the Coulomb-like QCD well,
the   $\eta_{c}^{\prime}$ and $\psi^{\prime}$
start to sample the confinement region of the QCD potential.  Most theoretical
estimates of $\Delta_{hf}^{\prime}$ are 
higher than
the new measurements\cite{Zaza}, perhaps because they tend to assume
a scalar QCD potential in the confinement region.  A recent lattice
result\cite{Okamoto}, 
which is quenched and thus ignores dynamical light quarks,
gives a span of values of $40 < \Delta_{hf}^{\prime} < 74$ MeV,
depending on input parameters.  This situation warrants a new look
at models and an unquenched lattice calculation.

\section{\bf A Sampling of $Q\overline{Q}$ Vector Results}

The vector $\psi$ and $\Upsilon$ resonances are directly produced
by $e^{+}e^{-}$ colliders via annihilation to a virtual photon.  With
large data sets recently obtained by CLEO at CESR and BES at BEPC,
it is not surprising that these two experiments dominate this
portion of the onia news.

\subsection{${\cal B}_{\mu\mu}$ for the $\Upsilon$ States}

Good measurements of ${\cal B}(\Upsilon(nS)\to\mu^{+}\mu^{-}) \equiv
{\cal B}_{\mu\mu}$(nS) are particularly important.  They play 
a crucial role in determining the resonance widths, in that
these states are significantly more narrow than the 
collider beam energy spread.
CLEO plans to measure $\Gamma_{ee}$ to within a few percent from
its careful scans of the resonant line shapes.  Learning $\Gamma_{tot}$
to $\sim 5\%$ thus means measuring ${\cal B}_{\mu\mu}$ to
similar precision. Also, ${\cal B}_{\mu\mu}$ is needed to
evaluate many other 
important branching fractions, in that
often what is really measured 
is the product 
${\cal B}(X \to Y\Upsilon$(nS)$)\cdot{\cal B}(\Upsilon$(nS)$\to\mu^{+}\mu^{-})$.
  
\begin{figure}[htb]
 \centerline{\hbox{\includegraphics[width=13cm]{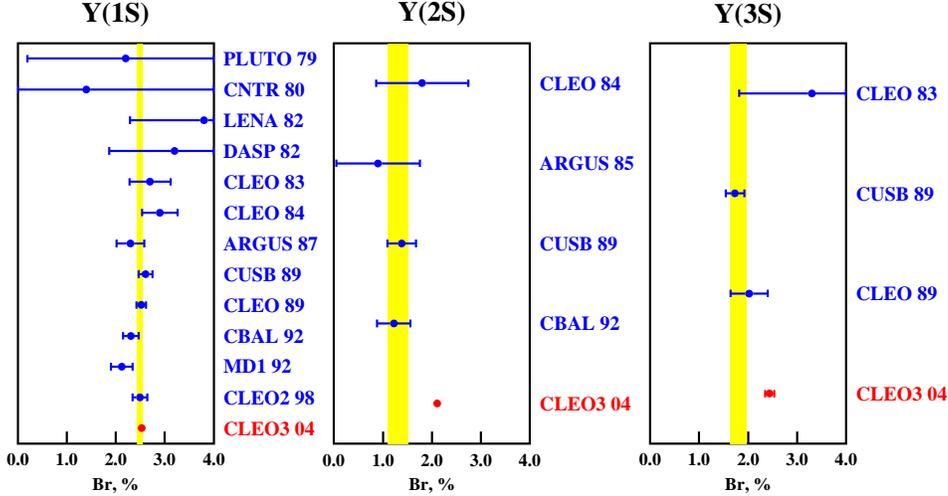}}}
 \caption{\it Comparison of the new CLEO results for $B_{\mu\mu}$
to prior measurements and, in the vertical stripe, the
current PDG world average.}
 \label{fig:Bmumu} 
\end{figure}

CLEO has both the statistical
power (over 5 million of each of the three bound-state $\Upsilon$
resonances) and the control of systematics to make such precise measurements
of ${\cal B}_{\mu\mu}$.  Their preliminary results\cite{Istvan}
are  ${\cal B}_{\mu\mu}$(nS) = 
$(2.53\pm 0.02\pm 0.05)\%$,
$(2.11\pm 0.03\pm 0.05)\%$, and
$(2.44\pm 0.07\pm 0.05)\%$ for n = 1, 2, 3, respectively.  For the
$\Upsilon$(1S) this agrees with the current PDG average\cite{PDG2002}, 
but the
results for the two radial excitations are significantly higher than 
in the PDG.  These results are shown graphically
in Fig.~\ref{fig:Bmumu}.

\subsection{From $Q\overline{Q}$ to $Q^{\prime}\overline{Q^{\prime}}$}

The {\it color octet} model was developed to help explain
large charmonium production rates at hadron colliders by having
$c\overline{c}$ produced from a single gluon (hence, as a color
{\it octet} state) and then perturbatively shedding a soft gluon before
emerging from the interaction.  
Models having two gluons form the $c\overline{c}$
meson as a final state particle are called {\it color singlet} 
models.  A good place to test such octet\cite{CKY} 
and singlet\cite{LXW} models is the
glue-rich environment of $\Upsilon$ decay.  While both predict
roughly equal rates for $\Upsilon\to J/\psi X$, the singlet momentum
spectrum is much softer in that a second $c$ and $\overline{c}$ must
hadronize as charmed mesons.


CLEO\cite{Blusk} has measured 
${\cal B}(\Upsilon\to J/\psi X) = (6.4 \pm 0.4 \pm 0.6) \times 10^{-4}$
which includes feed-down from other charmonia ($\psi^{\prime}$, ...);
this is consistent with the predictions of the models.  However the
measured $J/\psi$ momentum spectrum 
is much too
soft to support the color octet model.  Further the observation 
of $\Upsilon\to \psi^{\prime} X$, with
${\cal B}(\Upsilon\to \psi^{\prime}X) 
= (0.41 \pm 0.11 \pm 0.08) 
\times
{\cal B}(\Upsilon\to J/\psi X)$,
and the decay rates of $\Upsilon$ to $\chi_{c1}$ and $\chi_{c2}$ 
are above the octet model predictions.

\subsection{$\psi^{\prime}~v.~J/\psi$: The ``12\%'' Puzzle}

Major progress has been made on a long standing puzzle
in vector charmonium.  Because three-gluon decay to hadrons of
$c\overline{c}$ and electro-magnetic production/decay of
such states via a virtual photon both depend on the wave-function
overlap ($|\Psi (0)|^{2}$) one naively expects:

\begin{equation}
\label{eqn:rhopi}
\frac{{\cal B}(\psi^{\prime}\to h)}{{\cal B}(J/\psi\to h)} 
\equiv Q_{h}
= Q_{ee} 
\equiv 
\frac{{\cal B}(\psi^{\prime}\to e^{+}e^{-})}
{{\cal B}(J/\psi\to e^{+}e^{-})} \sim 12\%~.
\end{equation}

There are many complications, caveats and considerations
\footnote{For example,
the running of $\alpha_{s}$, form factor dependences on
$S$, helicity conservation issues, non-relativistic effects,
interference with continuum, ...}
to this ``equality'', so one should not be too surprised at
small deviations.  However, two particular hadronic modes,
$\rho\pi$ and $K^{*}\overline{K}$, both vector-pseudoscalar 
(V-P), are blatant
violators of Eqn.~\ref{eqn:rhopi}, with limits\cite{PDG2002} of
$Q_{h}/Q_{ee} < 0.1$.

\begin{figure}[htbp]
  \centerline{\hbox{ \hspace{0.2cm}
    \includegraphics[width=5.8cm]{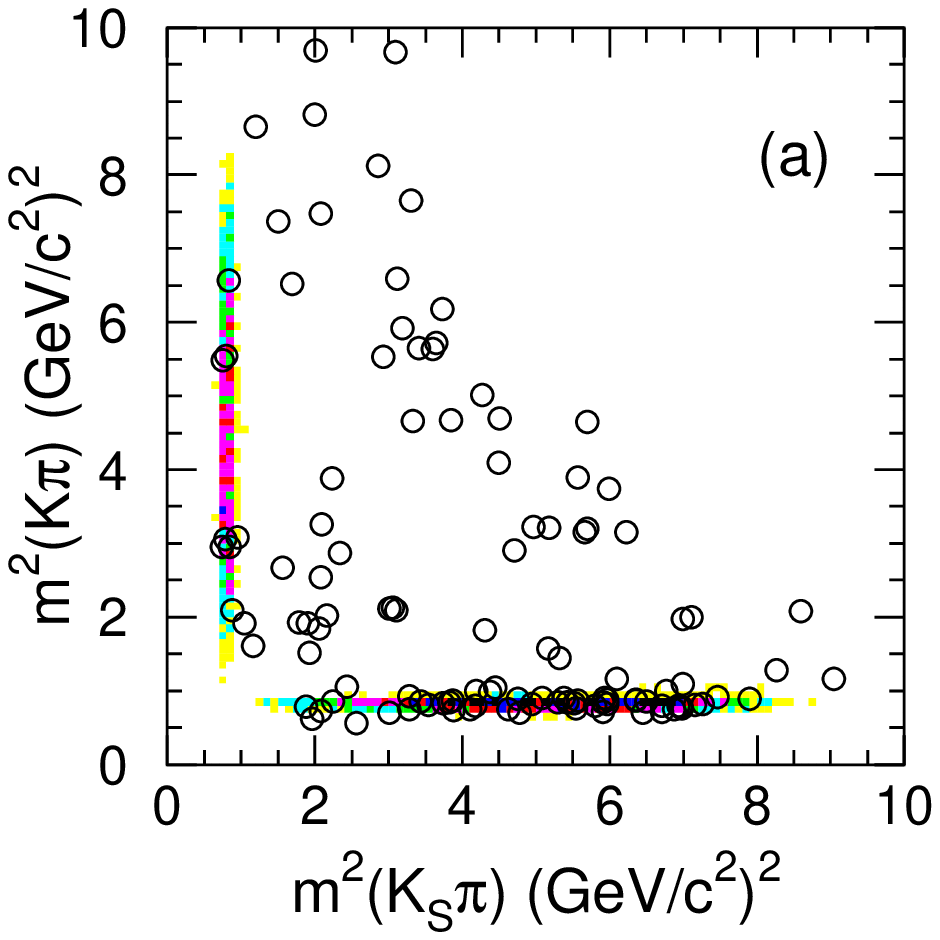}
    \hspace{0.3cm}
    \includegraphics[width=6.2cm]{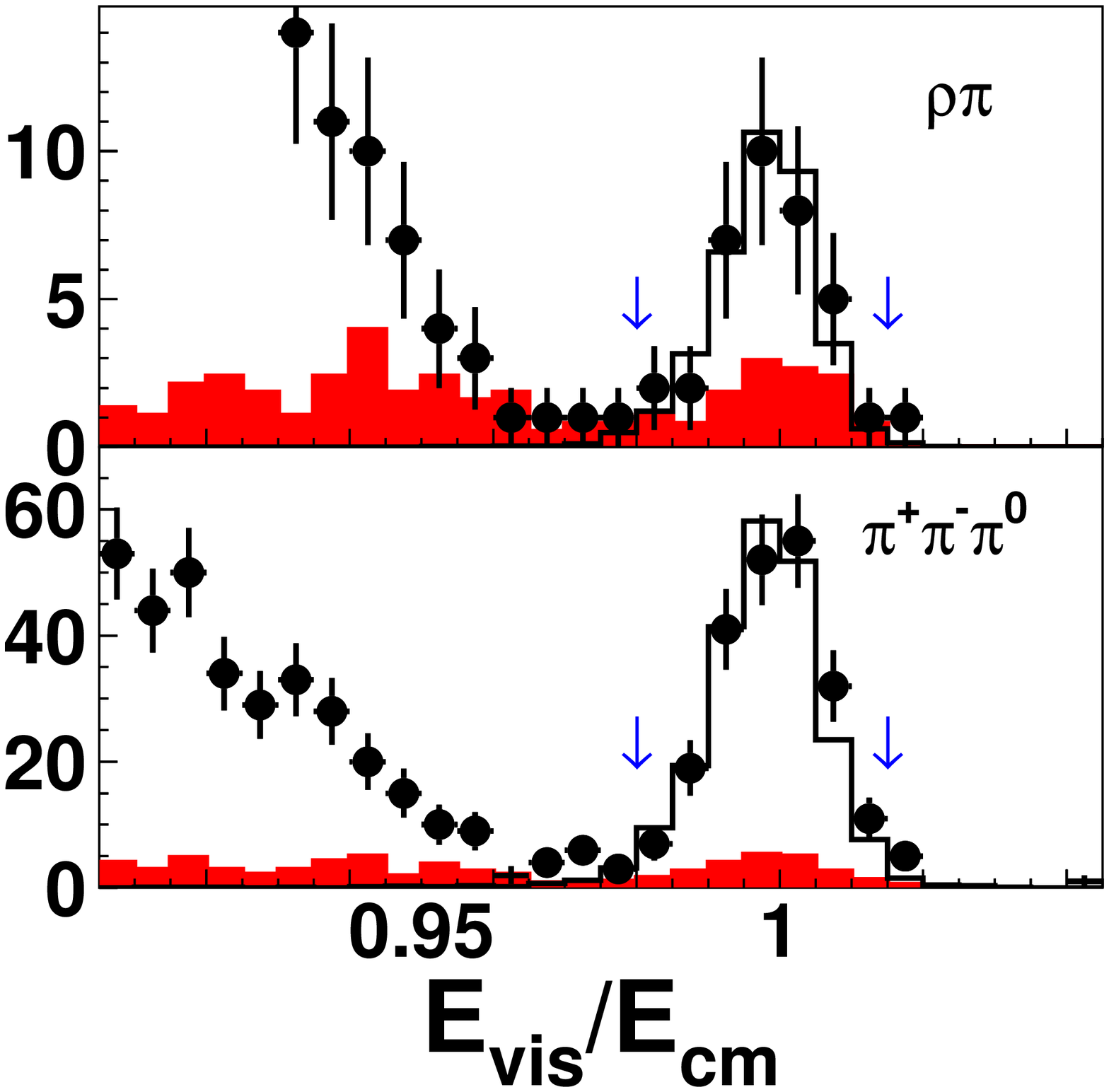}
    } 
   }  
  \caption{\it On the left is the BES Dalitz plot from their
analysis of $\psi^{\prime}\to K^{0}_{S}K^{\pm}\pi^{\mp}$; there
is a clear horizontal (vertical) enhancement corresponding to 
$K^{*0}K^{0}_{S}$ ($K^{*\mp}K^{\pm}$) decays. 
On the right is the CLEO distribution
of fractional visible energy in $\psi^{\prime}\to\pi^{+}\pi^{-}\pi^{0}$;
the upper of the two spectra has had a $\rho$ meson selected from the
associated Dalitz plot projections. }
 \label{fig:BESCLEOPV} 
\end{figure}

Big new data sets at BES/BEPC (14 million $\psi^{\prime}$ decays
and 6.4 pb$^{-1}$ of continuum) and CLEO-c/CESR-c
($\sim 3$ million $\psi^{\prime}$ decays from 5.5 pb$^{-1}$ of luminosity
and $\sim$ 20 pb$^{-1}$ of continuum) have come to rescue.
BES has presented evidence\cite{BESPV} for the modes
$K^{*0}\overline{K^{0}}$ and $K^{*+}K^{-}$ in the final state
$K^{0}_{S}K^{\pm}\pi^{\mp}$, as depicted in their Dalitz plot
in Fig.~\ref{fig:BESCLEOPV}.  There is a large signal in the
neutral mode, giving 
${\cal B}(\psi^{\prime}\to K^{*0}\overline{K^{0}}) = 
(15.0 \pm 2.1 \pm 1.7) \times 10^{-5}$ and
$Q_{h} = (3.6 \pm 0.7)\%$ (slight suppression relative to $J/\psi$).
The charged mode has a significant signal as well ($3.5\sigma$)
and gives
${\cal B}(\psi^{\prime}\to K^{*+}K^{-}) = 
(2.9 \pm 1.3 \pm 0.4) \times 10^{-5}$ and
$Q_{h} = (0.6 \pm 0.3)\%$ (heavily suppressed relative to $J/\psi$ $\!$).
Thus, there is a large difference in these isospin-related final
states.  The small continuum sample of BES does not allow for a subtraction
of this possible background source.

CLEO has presented\cite{CLEOPV} results for $Q_{h}/Q_{ee}$ for a
large number of modes, including, for the first time, $\rho\pi$
in the $\pi^{+}\pi^{-}\pi^{0}$ final state.  They have made a
continuum {\it subtraction}, but assume no interference contribution.
After projecting the
Dalitz plot to obtain samples of $\rho^{0}$ and $\rho^{\pm}$, they
extract the yield by demanding no missing energy, as shown to the
right of Fig.~\ref{fig:BESCLEOPV}.  Their results in the 
$\pi^{+}\pi^{-}\pi^{0}$ final state are 
$Q_{\rho\pi}/Q_{ee} = 0.016 \pm 0.006$ and
$Q_{\pi^{+}\pi^{-}\pi^{0}}/Q_{ee} = 0.053 \pm 0.011$.

\begin{figure}[htb]
 \centerline{\hbox{\includegraphics[width=10cm, height=9cm]{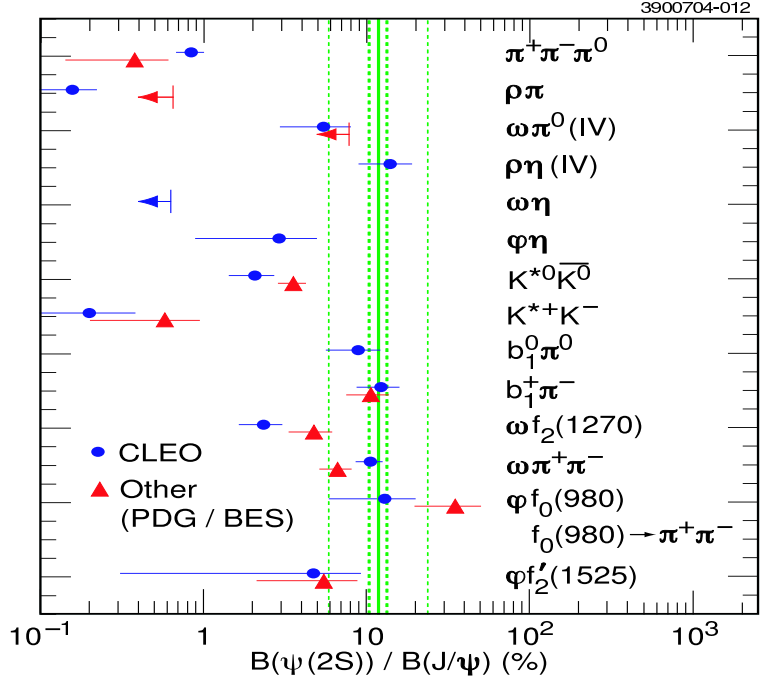}}}
 \caption{\it $Q_{h}$, the ratio of branching fractions
to specific hadronic states for the $\psi^{\prime} = \psi$(2S)
and $J/\psi$.  The solid vertical line is the naive expectation
for this ratio: $Q_{h} = Q_{ee} = 12\%$.  The inner dotted lines show
the present uncertainties in $Q_{ee}$; the outer dotted lines are
at $Q_{ee}/2$ and $2Q_{ee}$.}
 \label{fig:puzzle} 
\end{figure}

The status of this ``puzzle'' is represented pictorially in
Fig.~\ref{fig:puzzle}.
Some of the important features:
{\it (i)} the $\rho\pi$ and $K^{*}K$ modes have now been {\it measured};
{\it (ii)} the isospin violating (and hence, electro-magnetic) modes
seem to support the ``12\% rule''; {\it (iii)} the isospin-related
modes $K^{*0}\overline{K^{0}}$ and $K^{*+}K^{-}$ seem quite
different in their behavior;
{\it (iv)} the A-P states do not seem suppressed, while V-T states are
suppressed by about a factor of $\sim 5$ ...  not nearly as much
as the V-P states $\rho\pi$ and $K^{*+}K^{-}$.

An open question is whether the suppression is due to $S$-$D$
mixing, meaning a possible enhancement in V-P decays of the 
$\psi$(3770), by virtue of common virtual $D\overline{D}$ loops.

\subsection{$\psi^{\prime}~v.~J/\psi$: Relative Phases}

Another interesting comparison of the $J/\psi$ and $\psi^{\prime}$
wave-functions is the relative phase for each of these
to decay strongly via $c\overline{c}$ annihilation to 
three gluons (``S'')
as opposed to electro-magnetically via a single virtual photon (``EM'').
The decay $(c\overline{c}) \to K^{+}K^{-}$ can proceed by either
of these two routes, but $(c\overline{c}) \to K^{0}\overline{K^{0}}$
is purely a strong decay (SU(3) symmetry) and 
$(c\overline{c}) \to \pi^{+}\pi^{-}$ is purely electro-magnetic
(G-parity). Hence
\begin{equation}
\label{eqn:BES}
|{\cal A}(K^{+}K^{-})|^{2} =
|{\cal A}(\pi^{+}\pi^{-})|^{2} +
|{\cal A}(K^{0}\overline{K^{0}})|^{2} +
2\cdot |{\cal A}(\pi^{+}\pi^{-})|\cdot |{\cal A}(K^{0}\overline{K^{0}})| \cdot cos\phi_{S,EM} .
\end{equation}

Previously BES\cite{psiangle} 
has measured $cos\phi_{S,EM}(J/\psi) = (90 \pm 10)^{\circ}$.
With $14\times 10^{6}$ 
$\psi^{\prime}$ events they have observed a large
sample of mono-chromatic $K^{0}_{S}$ candidates from which they
obtain\cite{2BES} ${\cal B}(\psi^{\prime}\to K^{0}_{S}K^{0}_{L})$ =
$(5.24 \pm 0.47 \pm 0.48) \times 10^{-5}$.  Combining this with
various prior results on the $\pi^{+}\pi^{-}$ and $K^{+}K^{-}$
channels gives $cos\phi_{S,EM}(\psi^{\prime}) = (-89 \pm 29)^{\circ}$
or $(121 \pm 27)^{\circ}$.   Thus, $cos\phi_{S,EM}(\psi^{\prime})$ and
$cos\phi_{S,EM}(J/\psi)$ are consistent. 

BES also re-measured\cite{BESKSKL} ${\cal B}(J/\psi\to K^{0}_{S}K^{0}_{L})$ 
from their 55 million $J/\psi$ sample as
$(1.82 \pm 0.04 \pm 0.13) \times 10^{-4}$, meaning
${\cal B}(\psi^{\prime}\to K^{0}_{S}K^{0}_{L})$/
${\cal B}(J/\psi\to K^{0}_{S}K^{0}_{L})$
= $(29 \pm 4)\%$, somewhat large for the ``12\% rule'' 
represented by
Eqn.~\ref{eqn:rhopi}.


\section{\bf $Q\overline{Q}$ States with L = 2 (``D'')}

Bottomonium (see Fig.~\ref{fig:charmbottom}) is the only QCD
system with a stable member that has two units of orbital angular
momentum.  This makes these ``D'' states important to test
spin-orbit and high-L effects in both models and LQCD.  The CLEO
analysis of the four photon cascade
$\Upsilon$(3S)$\to \gamma_{1}\gamma_{2}$ ``D''
$\to\gamma_{1}\gamma_{2}\gamma_{3}\gamma_{4}\ell^{+}\ell^{-}$ 
is now final\cite{CLEOD}.  A single state is observed at
$M_{D} = (10161 \pm 0.6 \pm 1.6)$ MeV at $> 10\sigma$
significance.  The decay rate and intermediate $\chi_{b}^{\prime}$
and $\chi_{b}$ 
assignments are consistent with this being the $1^{3}D_{2}$ state.

\begin{figure}[htbp]
  \centerline{\hbox{ \hspace{0.2cm}
    \includegraphics[width=6.3cm]{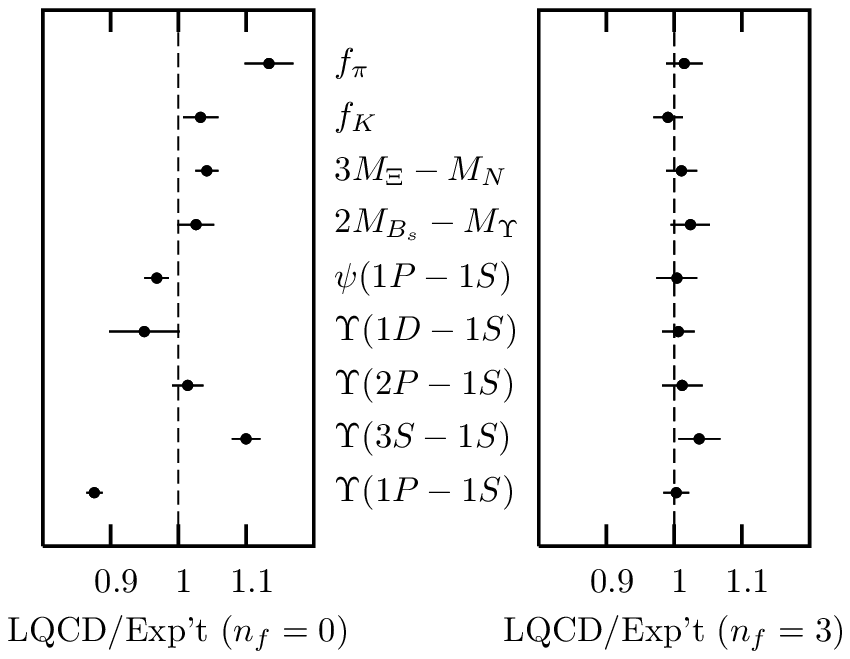}
    \hspace{0.3cm}
    \includegraphics[width=5.7cm,height=4.9cm]{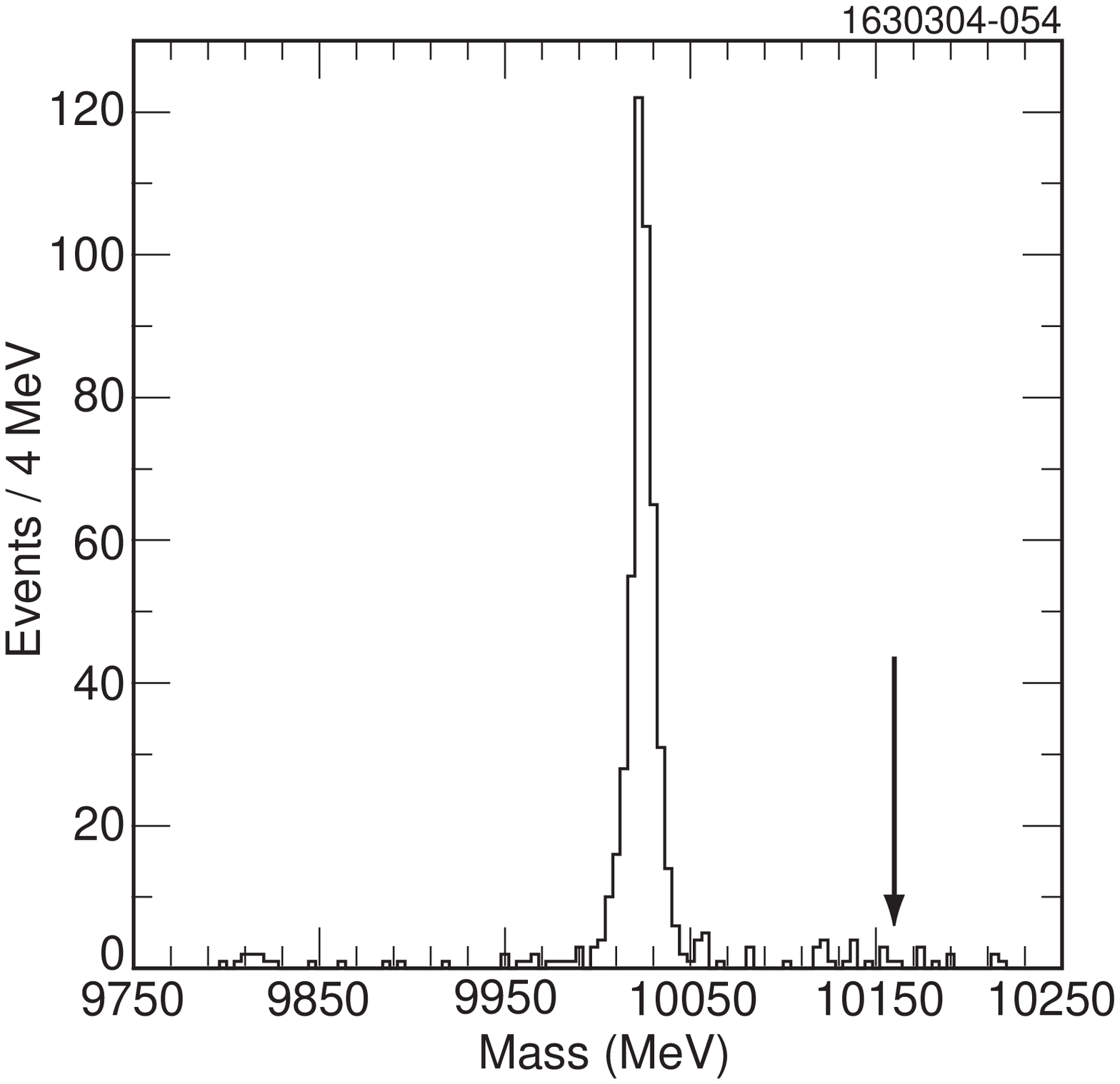}
    } 
   }  
  \caption{\it On the left is the current status of nine
``golden'' quantities by which to compare lattice QCD
calculations and experiment, with the most recent entry
being the ``D-S'' mass splitting in $b\overline{b}$.  To the
right is recoil mass against the two photons in
$\Upsilon$(3S)$\to\gamma\gamma[\pi^{+}\pi^{-}\Upsilon$(1S)$]$.
The observed peak is due to E1 cascade to the $\Upsilon$(2S);
no rate is observed for decay from the new $\Upsilon (1^{3}D_{2})$
state.}
 \label{fig:LQCD} 
\end{figure}

The impact of this result on LQCD\cite{Davies} 
is shown in the left portion of Fig.~\ref{fig:LQCD}.  Plotted first
results for nine ``golden'' quantities from
analyses 
that were ``quenched'', thus not incorporating the
effects of dynamical light quarks.  Next to this are recent
{\it un}quenched results with $uds$ quarks included.  None
of the nine quantities are used as ``input''.  These show LQCD being
able to attain better than 5\% accuracy in these quantities,
one of which is the D-S splitting in bottomonium.

Turning to charmonium, there is the continuing puzzle of the
nature of the $\psi$(3770), which one can characterize
as being $\alpha\cdot n^{3}S_{1} + \beta\cdot 1^{3}D_{1}
+ \gamma\cdot D\overline{D}$.   Clearly $\alpha \ne 0$ in that
the $\psi$(3770) {\it is} directly produced in $e^{+}e^{-}$
annihilation.  Belle\cite{Belle3770} has recently observed 
a solid $\psi$(3770) signal in $B^{+} \to K^{+}D^{0}\overline{D^{0}}$,
from which they extract
${\cal B}(B \to K \psi{\rm (3770)}) 
= (4.8 \pm 1.1 \pm 0.7) \times 10^{-4}
\sim 2/3~{\cal B}(B \to K \psi^{\prime})$.  
This comparison would imply that there is large S-D mixing,
although color-octet models might be able to accommodate this
rate for a pure ``D'' state ($\beta \sim 1$).

An important mode to investigate would seem to be 
$\psi$(3770)$\to\pi^{+}\pi^{-}J/\psi$.  A compilation of
MarkII and BES results\cite{Rosner3770} gives
$\Gamma(\pi^{+}\pi^{-}J/\psi) = (43 \pm 14)$ keV, which
is near the {\it upper bound} for this rate as analyzed by
CLEO\cite{TS}. The Kuang-Yan prediction\cite{KY} for this rate
is 20-107 keV, depending on the level of S-D mixing; however,
the CLEO limit\cite{CLEOD}, depicted in Fig.~\ref{fig:LQCD},
for $\Upsilon$(1D)$\to\pi^{+}\pi^{-}J/\psi$,
is some seven times below\cite{RosnerD} that predicted for a
such a
D state in the Kuang-Yan model, casting doubt on whether
the $\Gamma(\pi^{+}\pi^{-}J/\psi)$ measured for
the $\psi$(3770) is {\it really}
consistent with theoretical expectations.  Once the decay to
$\pi^{+}\pi^{-}J/\psi$ is solidly
observed, the angular distributions of the decay products should
help sort out\cite{Voloshin} $\alpha, \beta, \gamma$.

\section{\bf $Q\overline{Q}$ States with L = 1 (``$\chi$'')}

Although copiously produced in E1 transitions
from the vector mesons directly obtained in 
$e^{+}e^{-}$ annihilation, rather little is known about
these $\chi_{c}$ and $\chi_{b}$ states.  

CLEO has finalized\cite{Pedlar} its analysis of
$\chi_{bJ}^{\prime} \to \omega\Upsilon$(1S) for 
$J=1,2$.  There is so little $Q$ value for this decay
that the process is kinematically forbidden for J=0!
Nonetheless, the quoted branching fractions are large:
${\cal B}(\chi_{b1}^{\prime} \to \omega\Upsilon{\rm (1S)})
= (1.6 \pm 0.3 \pm 0.2)\%$
and
${\cal B}(\chi_{b2}^{\prime} \to \omega\Upsilon{\rm (1S)})
= (1.1 \pm 0.3 \pm 0.1)\%$, to be compared to the
$\sim 7 \%$ branching fraction for the E1 photon transitions
$\chi_{bJ}^{\prime}\to\gamma\Upsilon$(1S).
The rates for this $\omega$ transition
are therefore roughly equal for $J=1$ and $J=2$, 
as predicted\cite{Voloshin}.

The BES collaboration has used its $\chi_{c}$ sample to test
the color-octet model in baryonic decays.  The color singlet
model is unable to generate the observed rates for, {\it e.g.,}
$\chi_{cJ}\to p \overline{p}$.  Adding in color octet
contributions\cite{Wong} predicts that the ratio
${\cal R}_{B} = \Gamma(\Lambda\overline{\Lambda})/\Gamma(p\overline{p})$
is $\sim 0.60$ and $\sim 0.45$ for $\chi_{c1}$ and $\chi_{c2}$ decays,
respectively.
The BES collaboration \cite{BESchi} 
last year reported on the $\Lambda\overline{\Lambda}$
channel; now they have very clean signals for all three $J$ states
decaying into $p \overline{p}$.  These lead to experimental values
of ${\cal R}_{B} = 4.6\pm 2.3$ and $5.1\pm 3.1$ for $J=1$ and $J=2$.
While the uncertainties are large, they nonetheless tend to show
the $\Lambda\overline{\Lambda}$ 
channel enhanced, not suppressed, with respect to
$p \overline{p}$.

\section{\bf The $X$(3872) - New Quarkonium-like State !!}

\begin{figure}[htbp]
  \centerline{\hbox{ \hspace{0.2cm}
    \includegraphics[width=6.0cm]{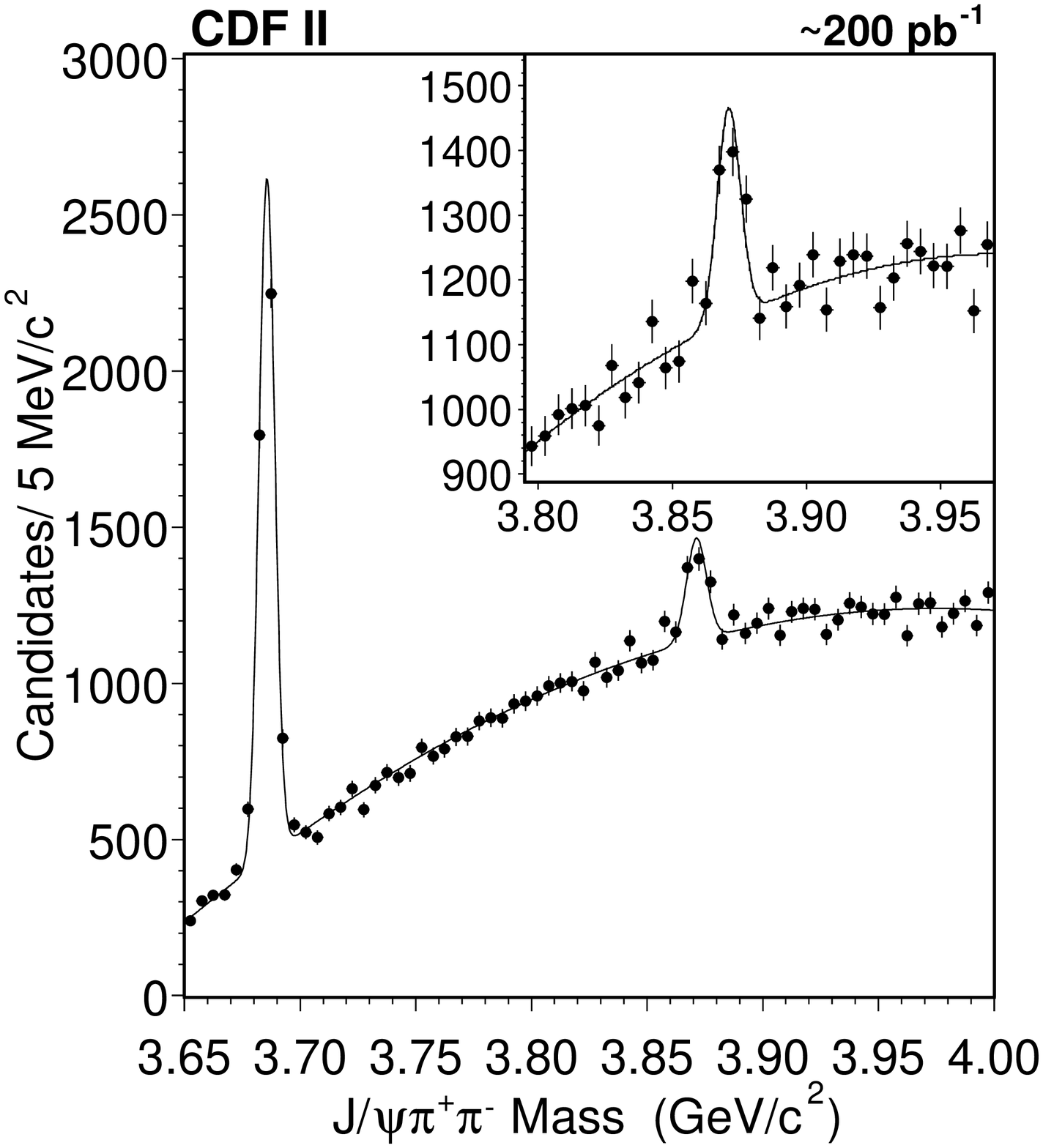}
    \hspace{0.3cm}
    \includegraphics[width=6.0cm]{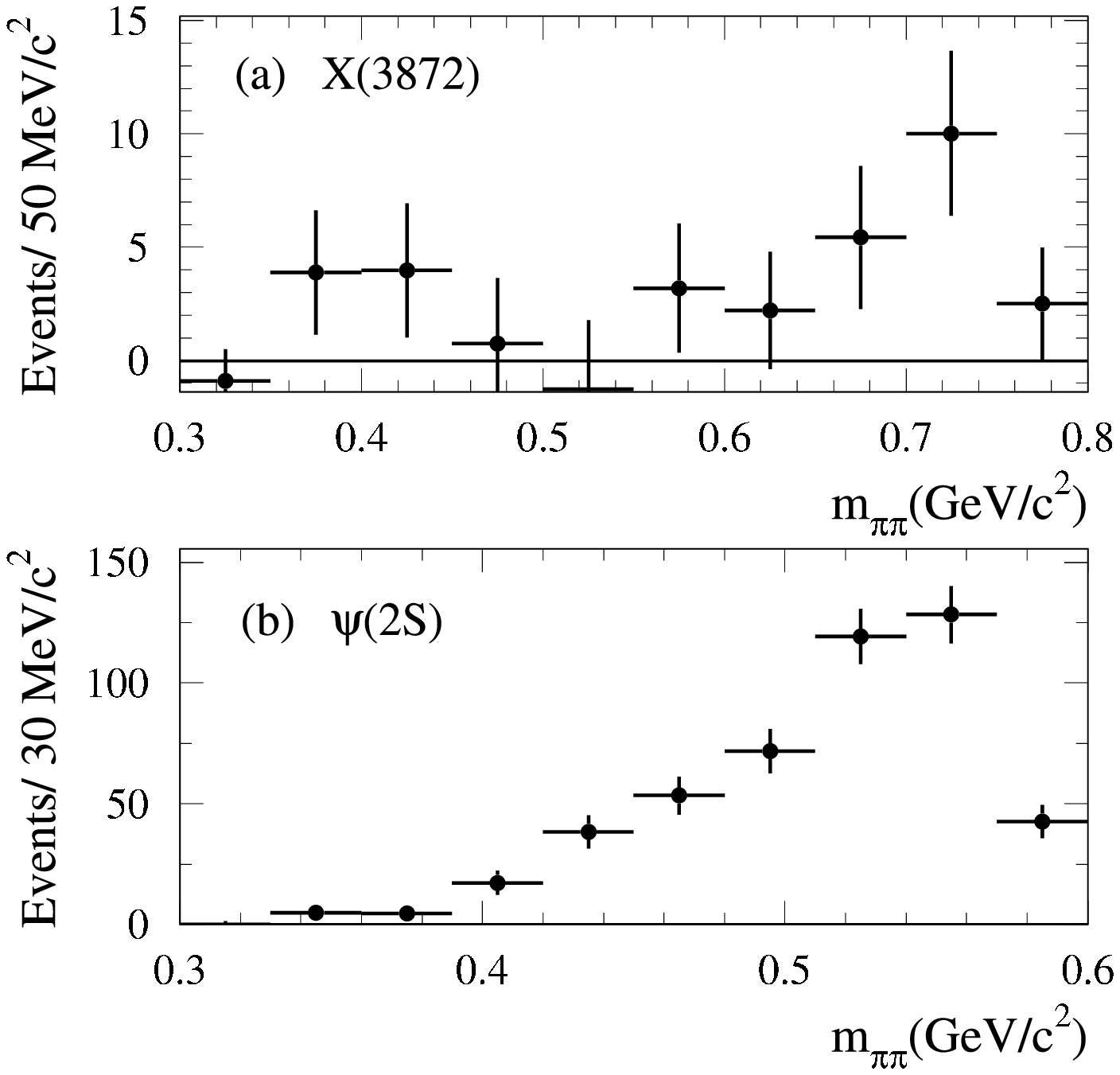}
    } 
   }  
  \caption{\it At QWGII, CDF confirmed the Belle observation of
the X(3872), showing the left plot, with clear evidence
of the $\psi^{\prime}$ and the X ($>11\sigma$ significance);
the events were required to have a di-pion invariant mass in excess
of 500 MeV.  To the right are the BaBar spectra of this di-pion mass
for both the X and $\psi^{\prime}$ regions. }
 \label{fig:X3872} 
\end{figure}

The hottest onia news at last year's Lepton-Photon 
meeting\cite{TS} was the observation ($>10\sigma$) 
by Belle\cite{Belle3872}
of a new ``charmonium-like'' state at 3872 MeV produced in 
$B^{+} \to K^{+} X$
and decaying to $J/\psi~\pi^{+}\pi^{-}$.  Since then, it has
been seen clearly by CDF\cite{CDF3872} in hadro-production
(as reported at 
QWGII\cite{QWG} and shown in Fig.~\ref{fig:X3872})
as well as by D0\cite{D03872} and BaBar\cite{BaBar3872}.

Of particularly note is the nature of the $\pi\pi$ system
in the $X$ decay.  Belle\cite{Belle3872} reported a dipion mass
($m_{\pi\pi}$) consistent with the decay $X \to \rho~J/\psi$,
which would
imply that $C_{X} = +1$.  This tendency toward 
large $m_{\pi\pi}$ was actually
{\it used} by CDF and D0 to enhance their signals. 
Fig.~\ref{fig:X3872} also shows the $m_{\pi\pi}$ spectrum 
as reported
by BaBar\cite{BaBar3872} for the $X$ and for their large
$\psi^{\prime}$(3686) sample.  Is the high-mass region {\it really} a
$\rho$ or is it just a mimic of the similar shape in
$\psi^{\prime}$ decay?  Is there a second, low-mass peak in the
$m_{\pi\pi}$ spectrum, similar to what has been observed\cite{3S1Spipi} 
in  $\Upsilon$(3S)$\to\pi\pi\Upsilon$(1S)?  Do experiments see the more
difficult channel $X\to\pi^{0}\pi^{0}$, which would imply
$C_{X} = -1$?  One should keep watching this di-pion system
as the sample sizes increase with more data at the Tevatron and
the $B$-factories.

D0\cite{D03872} has looked at six aspects of the production
and decay of the $X$(3872) as compared to the $\psi^{\prime}$(3686):
$p_{T}$ to the jet, range of rapidity, isolation, decay length
distribution, helicity angles of the muons and of the pions.  In all
cases the $X$ looks like the well-established $c\overline{c}$ state.

The four-experiment average has $M_{X} = 3872.2 \pm 0.5$ MeV,
tantalizingly close to the mass of a $D^{0}D^{0*}$ pair
at\cite{PDG2002} $3871.2\pm 0.9$ MeV but significantly below
the corresponding charged system, with a mass of a
$D^{+}D^{-*}$ pair being $3879.3 \pm 1.0$ MeV.  This ``coincidence'',
coupled with the fact\cite{Belle3872, BaBar3872} that
\begin{equation}
\frac{
{\cal B}(B^{+}\to K^{+}X{\rm (3872)})\cdot{\cal B}(X\to J/\psi~\pi^{+}\pi^{-})
}
{
{\cal B}(B^{+}\to K^{+}\psi^{\prime}{\rm (3686)})\cdot{\cal B}(\psi^{\prime}\to J/\psi~\pi^{+}\pi^{-})
}
= 0.062 \pm 0.011~,
\label{eqn:X3872}
\end{equation}
leads to the conjecture that the $X$(3872) has a $c\overline{c}$ ``core'' and
a large $D^{0}D^{0*}$ ``molecular'' component.

There has been a lot of activity (with a lot more ongoing!) to determine
the quantum numbers
of this new state.  Searches by Belle\cite{Belle3872} for the
decays $\gamma\chi_{c1}$, $\gamma\chi_{c2}$, and
$\gamma J/\psi$ have shown it unlikely that $X$(3872) is 
$1^{3}D_{2}$, $1^{3}D_{3}$, or $\chi_{c}^{\prime} (2^{3}P_{J})$, 
respectively. Other studies of BES and CLEO data\cite{JPC}
in radiative return
and two-photon fusion production
tend to disfavor vector states ($J^{PC}=1^{--}$) and those with
even $J$ and positive $C$-parity.  

This is certainly a most curious state and hopefully much more
will be known about it by PIC2005!
 
\section{\bf Summary and Acknowledgments}

As is evident, heavy quarkonia continue to
be a source of great energy and excitement, with
many recent advances and many open questions:
a stable ``D'' state has been firmly established and
agrees nicely with the LQCD prediction; the
long-awaited $\eta^{\prime}_{c}$ has also been
firmly established with a small hyper-fine mass
splitting to the $\psi^{\prime}$; several analyses are
confronting the color octet production model; V-P states
have finally been observed in $\psi^{\prime}$ decay with
suppressions relative to $J/\psi$ decay of roughly a factor
of 50; while more is being learned about the nature of
the $\psi$(3770), it still remains a puzzle; even more
of a mystery is the very narrow state at 3872 MeV!  Lots
yet to do!  

I wish to thank all in the BES, BaBar, Belle, CDF, CLEO
and D0 groups who helped me put together this attempt at
a selective summary.
  


\end{document}